\documentclass[twocolumn,epj]{webofc}

\usepackage[varg]{txfonts}   
\usepackage{amsmath,amssymb,amsfonts}
\usepackage{graphicx}
\usepackage{booktabs,colortbl,multirow}
\usepackage{colordvi,color}

\newcommand{\be}{\begin{equation}} 
\newcommand{\ee}{\end{equation}} 

\newcommand{\units}[1]{~\mathrm{#1}}

\newcommand{\ctoprule}{\toprule[0.5mm]}
\newcommand{\cbottomrule}{\bottomrule[0.5mm]}

\newcommand{\hc}{\mathrm{h.c.}}

\woctitle{LHCP 2013}
\begin{document}
\title{Electroweak limits on physics beyond the Standard Model}
\author{Jorge de Blas\inst{1}\fnsep\thanks{\email{jdeblasm@nd.edu}}}
\institute{Department of Physics, University of Notre Dame, Notre Dame, IN 46556, USA}
\abstract{%
We briefly review the global Standard Model fit to electroweak precision data, and discuss the status of electroweak constraints on new interactions. We follow a general effective Lagrangian approach to obtain model-independent limits on the dimension-six operators, as well as on several common new physics extensions.
}
\maketitle
%


\section{Introduction}
\label{section_Intro}

The measurement of the Higgs mass at the LHC \cite{Chatrchyan:2012ufa,Aad:2012tfa} provides a direct determination of the last of the
Standard Model (SM) input parameters. With this measurement the global SM fit to Electroweak
Precision Data (EWPD) is overconstrained. At the minimum, the best fit values for the SM inputs are given by
\begin{table}[h]
\centering
\begin{tabular}{l l}
\!$M_H=125.7\pm0.42 \units{GeV}$&\!\!\!\!$[99^{+26}_{-21}\units{GeV}]$,\\[0.1cm]
\!$m_t=173.5 \pm 0.81 \units{GeV}$&\!\!\!\!$[175.5\pm2.2\units{GeV}]$,\\[0.1cm]
\!$M_Z=91.1879 \pm 0.0021 \units{GeV}$&\!\!\!\!$[91.197\pm0.010\units{GeV}]$,\\[0.1cm]
\!$\alpha_s(M_Z^2)=0.1186 \pm 0.0007$&\!\!\!\!$[0.1187\pm0.0026]$,\\[0.1cm]
\!$\Delta \alpha_{\mbox{\scriptsize had}}^{(5)}(M_Z^2)=(2755 \pm 10)\cdot10^{-5}$&\!\!\!\!$[(2735 \pm 41)\cdot10^{-5}]$.\\
\end{tabular}
\end{table}

The results in brackets are the values obtained by repeating the fit after excluding each direct determination from the data. Thus, the bracketed numbers represent the indirect determination of the corresponding input parameter from EWPD. As can be seen, all indirect determinations are around or below $1~\sigma$ from the direct measurements, illustrating the consistency between data and the SM. More precisely, the goodness of the fit can be quantified by the $\chi^2/n_{\mbox{\scriptsize d.o.f.}}=1.05$, which corresponds to a reasonably good probability: $p\mbox{-value}=0.26$. Our electroweak fits include the usual $Z$-pole data \cite{ALEPH:2005ab}, $\Delta \alpha_{\mbox{\scriptsize had}}^{(5)}(M_Z^2)$ \cite{Davier:2010nc}, the latest values for $\alpha_s(M_Z^2)$ \cite{Pich:2013sqa}, the top mass \cite{CDF:2013jga} and the $W$ mass and width \cite{Group:2012gb}, as well as several low-energy measurements \cite{Beringer:1900zz,Erler:2013xha}. We also include the final results of $e^+ e^-\rightarrow \bar{f} f$ from LEP 2 \cite{Schael:2013ita}.

Despite the general good agreement between the data and the SM predictions, when looking at the agreement for individual observables, there are a few significant discrepancies. It is also noteworthy that some of these appear to be centered in the bottom sector. First, we have the longstanding anomaly in the forward-backward asymmetry for the $b$ quark, $A_{FB}^b$, which is around $-2.5~\sigma$ away from the SM prediction. Second, the recent computation of the 2-loop corrections to $R_b$ results in a contribution of size similar to the experimental error \cite{Freitas:2012sy}. This moves the observable from a previous good agreement ($0.7~\sigma$) to a discrepancy of about $2.4~\sigma$, enhancing the tension in the bottom sector.

From a theoretical and ``aesthetic" point of view, on the other hand, it could be argued that the SM does not perform as well. It has too many free parameters and it is unable to provide answers to several fundamental questions. Moreover, it also suffers from naturalness/fine tuning problems, such as the hierarchy problem. These reasons suggest that the SM can only be realized as a low-energy approximation of a more fundamental theory. Naturalness arguments also lead to the conclusion that new physics beyond the SM should be present not far from the TeV scale. In this regard, once all the SM inputs are set, the existence of any extra contributions to electroweak precision observables can only be attributed to new physics. Hence, the predictive power of the electroweak fit can also be used to constrain the size of these new physics effects.


\section{Model-independent bounds on new physics}
\label{section_MI_NP}

In the absence of any hint about the nature of any possible physics beyond the SM, it is convenient to use a general model-independent description of new physics effects. This is provided by using an effective Lagrangian expansion,
\be
\begin{split}
&{\cal L}_{\mathrm{Eff}}=\sum_{d=4}^\infty\frac{1}{\Lambda^{d-4}}{\cal L}_d={\cal L}_{\mathrm{SM}}+\frac{1}{\Lambda}{\cal L}_5+\frac{1}{\Lambda^2}{\cal L}_6+\cdots,\\[0.1cm]
&\hspace{1.5cm}~{\cal L}_{d}=\sum_i \alpha_i^d {\cal O}_i^d,~~~~\left[{\cal O}_i^d\right]=d,
\label{EffL}
\end{split}
\ee
which parametrizes new physics effects at energies below the new physics scale, $\Lambda$, as contributions to the coefficients of the higher-dimensional operators in (\ref{EffL}). Assuming $\Lambda\gtrsim1\units{TeV}$, and taking into account the precision of current data, it suffices for most purposes to consider effects up to dimension six in the effective Lagrangian expansion. At this order, all possible operators have already been classified \cite{Buchmuller:1985jz,Grzadkowski:2010es}. The EWPD limits on dimension-six interactions that can be generated at the tree level upon integration of any extra scalars, fermions or vector bosons, and that can interfere with the SM amplitudes, were computed in \cite{delAguila:2011zs}. In Table \ref{table: indivOfit} we recompute those bounds including the most recent updates in the experimental data and theoretical predictions. As in \cite{delAguila:2011zs}, and unless otherwise is stated, we assume only one operator at a time in the fits, as well as flavor-diagonal and family-universal fermionic interactions whenever this applies. 

The updates in the fit compared to \cite{delAguila:2011zs} have several implications. First, since the operators ${\cal O}_{\phi}^{(3)}=(\phi^\dagger D_\mu \phi)((D^\mu\phi)^\dagger \phi)$ and ${\cal O}_{WB}=(\phi^\dagger \sigma_a\phi)W_{\mu\nu}^a B^{\mu\nu}$ (equivalent to the $T$ and $S$ oblique parameters, respectively) are strongly correlated with the Higgs mass, the precise measurement of the latter has a significant impact in the individual fit to each of these operators. (For the case of ${\cal O}_{\phi}^{(3)}$ [${\cal O}_{WB}$] including the measurement of $M_H$ reduces the size of the confidence interval by a factor $\sim 5~[\sim 3]$.)

Second, there are several updates in the data taken away from the $Z$ pole. These are present both in the low-energy observables and in the $e^+ e^-\rightarrow \bar{f}f$ data above the $Z$ pole from LEP 2. For instance, the measurement of the weak charge of $^{133}_{\hspace{0.125cm}55}$Cs in atomic parity violation experiments has moved from almost a perfect agreement with the SM to about $1.3~\sigma$ \cite{Dzuba:2012kx}. The final results from LEP 2 are all in good agreement with the SM. In particular, the small excess found in the hadronic cross section in the preliminary results has been reduced to $1.2~\sigma$. The new LEP 2 data also include determinations of the differential cross-sections in $e^+ e^-\rightarrow \mu^+\mu^-,\tau^+\tau^-$, not present in the analysis of \cite{delAguila:2011zs}. All these (low-energy and high-energy) observables are particularly relevant to constrain new physics in the form of four-fermion interactions, since the $Z$-pole data are mostly insensitive to such effects. Since most of the four-fermion operators only contribute to very few observables, a change in one particular observable can have a large impact in the confidence interval for the corresponding interaction. This explains the differences in some of the results compared to \cite{delAguila:2011zs}. On the other hand, the dimension-six operators $(\phi^\dagger i D_\mu \phi)(\overline{\psi}\gamma^\mu \psi)$ and $(\phi^\dagger i\sigma_a D_\mu \phi)(\overline{\psi}\gamma^\mu\sigma_a \psi)$ induce direct contributions to the $Z$-pole observables. Therefore, the changes and new additions of the non-$Z$ pole data have a smaller effect in many cases, especially for the leptonic operators. 

\begin{table}[th]
\caption{$95\%$ C.L. EWPD limits on the 
dimension-six operator coefficients. The limits are obtained 
from a fit considering only one operator at a time. Fermionic interactions are assumed 
to be flavor diagonal and family universal.  
\label{table: indivOfit}}
\centering
\begin{tabular*}{\columnwidth}{@{\extracolsep{\fill}} c c  c }
\ctoprule
Operator&\!\!\!\!\!\!\!\!\!\!\!Coefficient&\!\!$95\%$ C.L. EWPD limit\\
${\cal O}_i$&\!\!\!\!\!\!\!\!\!\!\!$\frac{\alpha_i}{\Lambda^2}$& [TeV$^{-2}$]\\[0.1cm]
\midrule
$\frac 12 (\overline{l_L}\gamma^\mu l_L)(\overline{l_L}\gamma_\mu l_L)$
&\!\!\!\!\!\!\!\!\!\!\!$\frac{\alpha_{ll}^{(1)}}{\Lambda^2}$&$\left[-0.058,0.037\right]$ \\[0.1cm]
$\frac 12 (\overline{l_L}\gamma^\mu\sigma_a l_L)(\overline{l_L}\gamma_\mu\sigma_a l_L)$&\!\!\!\!\!\!\!\!\!\!\!$\frac{\alpha_{ll}^{(3)}}{\Lambda^2}$&$\left[-0.007,0.008\right]$ \\[0.1cm]
$(\overline{l_L}\gamma^\mu l_L)(\overline{q_L}\gamma_\mu q_L)$&\!\!\!\!\!\!\!\!\!\!\!$\frac{\alpha_{lq}^{(1)}}{\Lambda^2}$&$\left[-0.012,0.055\right]$ \\[0.1cm]
$(\overline{l_L}\gamma^\mu\sigma_a l_L)(\overline{q_L}\gamma_\mu\sigma_a q_L)$&\!\!\!\!\!\!\!\!\!\!\!$\frac{\alpha_{lq}^{(3)}}{\Lambda^2}$&$\left[-0.006,0.012\right]$ \\[0.1cm]
\midrule
$\frac 12 (\overline{e_R}\gamma^\mu e_R)(\overline{e_R}\gamma_\mu e_R)$
&\!\!\!\!\!\!\!\!\!\!\!$\frac{\alpha_{ee}}{\Lambda^2}$&$\left[-0.051,0.009\right]$ \\[0.1cm]
$(\overline{e_R}\gamma^\mu e_R)(\overline{u_R}\gamma_\mu u_R)$&\!\!\!\!\!\!\!\!\!\!\!$\frac{\alpha_{eu}}{\Lambda^2}$&$\left[-0.097,0.017\right]$ \\[0.1cm]
$(\overline{e_R}\gamma^\mu e_R)(\overline{d_R}\gamma_\mu d_R)$&\!\!\!\!\!\!\!\!\!\!\!$\frac{\alpha_{ed}}{\Lambda^2}$&$\left[-0.077,0.040\right]$ \\[0.1cm]
\midrule
$(\overline{l_L} e_R)(\overline{e_R} l_L)$
&\!\!\!\!\!\!\!\!\!\!\!$\frac{\alpha_{le}}{\Lambda^2}$&$\left[-0.050,0.074\right]$ \\[0.1cm]
$(\overline{l_L} u_R)(\overline{u_R} l_L)$&\!\!\!\!\!\!\!\!\!\!\!$\frac{\alpha_{lu}}{\Lambda^2}$&$\left[-0.191,0.082\right]$ \\[0.1cm]
$(\overline{l_L} d_R)(\overline{d_R} l_L)$&\!\!\!\!\!\!\!\!\!\!\!$\frac{\alpha_{ld}}{\Lambda^2}$&$\left[-0.213,0.041\right]$ \\[0.1cm]
$(\overline{q_L} e_R)(\overline{e_R} q_L)$&\!\!\!\!\!\!\!\!\!\!\!$\frac{\alpha_{qe}}{\Lambda^2}$&$\left[-0.022,0.110\right]$ \\[0.1cm]
\midrule
$(\phi^\dagger iD_\mu \phi)(\overline{l_L}\gamma^\mu l_L)$
&\!\!\!\!\!\!\!\!\!\!\!$\frac{\alpha_{\phi l}^{(1)}}{\Lambda^2}$&$\left[-0.005,0.011\right]$ \\[0.1cm]
$(\phi^\dagger iD_\mu \phi)(\overline{q_L}\gamma^\mu q_L)$&\!\!\!\!\!\!\!\!\!\!\!$\frac{\alpha_{\phi q}^{(1)}}{\Lambda^2}$&$\left[-0.021,0.025\right]$ \\[0.1cm]
$(\phi^\dagger iD_\mu \phi)(\overline{e_R}\gamma^\mu e_R)$&\!\!\!\!\!\!\!\!\!\!\!$\frac{\alpha_{\phi e}^{(1)}}{\Lambda^2}$&$\left[-0.013,0.007\right]$ \\[0.1cm]
$(\phi^\dagger iD_\mu \phi)(\overline{u_R}\gamma^\mu u_R)$&\!\!\!\!\!\!\!\!\!\!\!$\frac{\alpha_{\phi u}^{(1)}}{\Lambda^2}$&$\left[-0.070,0.031\right]$ \\[0.1cm]
$(\phi^\dagger iD_\mu \phi)(\overline{d_R}\gamma^\mu d_R)$&\!\!\!\!\!\!\!\!\!\!\!$\frac{\alpha_{\phi d}^{(1)}}{\Lambda^2}$&$\left[-0.098,0.008\right]$ \\[0.1cm]
$(\phi^\dagger i\sigma_a D_\mu \phi)(\overline{l_L}\gamma^\mu\sigma_a l_L)$&\!\!\!\!\!\!\!\!\!\!\!$\frac{\alpha_{\phi l}^{(3)}}{\Lambda^2}$&$\left[-0.007,0.004\right]$ \\[0.1cm]
$(\phi^\dagger i\sigma_a D_\mu \phi)(\overline{q_L}\gamma^\mu\sigma_a q_L)$&\!\!\!\!\!\!\!\!\!\!\!$\frac{\alpha_{\phi q}^{(3)}}{\Lambda^2}$&$\left[-0.007,0.007\right]$ \\[0.1cm]
$(\phi^T i\sigma_2 D_\mu \phi)(\overline{u_R}\gamma^\mu d_R)$&\!\!\!\!\!\!\!\!\!\!\!$\frac{\alpha_{\phi ud}}{\Lambda^2}$&$\left[-0.017,0.022\right]$ \\[0.1cm]
\midrule
$(\phi^\dagger D_\mu \phi)((D^\mu\phi)^\dagger \phi)$
&\!\!\!\!\!\!\!\!\!\!\!$\frac{\alpha_{\phi}^{(3)}}{\Lambda^2}$&$\left[-0.023,0.006\right]$ \\[0.1cm]
$(\phi^\dagger \sigma_a\phi)W_{\mu\nu}^a B^{\mu\nu}$&\!\!\!\!\!\!\!\!\!\!\!$\frac{\alpha_{WB}}{\Lambda^2}$&$\left[-0.007,0.003\right]$ \\[0.1cm]
\cbottomrule
\end{tabular*}
\end{table}

It is interesting to examine what operators can loose the tension between theory and data for those observables that show a discrepancy with the SM predictions. For the specific case of $A_{FB}^b$ and $R_b$ measured at the $Z$ pole, this can be done introducing direct corrections to the neutral current bottom couplings, 
\be
\begin{split}
g_{L,R}^b=g_{L,R}^{b~\!\mathrm{SM}}+\delta g_{L,R}^b\hspace{1.75cm}\\[0.1cm]
(g_L^{b~\!\mathrm{SM}}=-\frac12+\frac 13\sin^2{\theta_W}, ~g_R^{b~\!\mathrm{SM}}=\frac 13\sin^2{\theta_W}).\nonumber
\end{split}
\ee
\vspace{0cm}

\begin{figure}[t]
\centering
\input{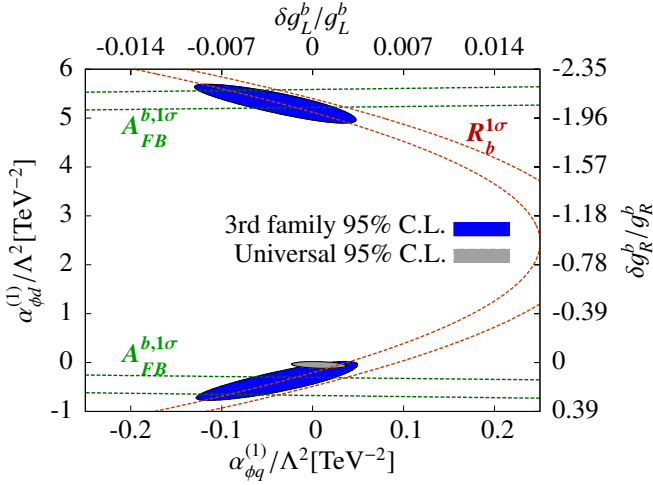}
\caption{Solutions to the $R_b$ and $A_{FB}^b$ anomalies within a SM extension with the dimension-six operators ${\cal O}_{\phi q}^{(1)}$ and ${\cal O}_{\phi d}^{(1)}$. The small (grey) ellipse represents the $95\%$ confidence region from a fit assuming diagonal and universal couplings, and it is unable to reconcile both observables with the experimental values. The large (blue) $95\%$ C.L. ellipses represent the two solutions obtained from a fit assuming interactions only with the third family.
\label{Fig_Zbb}}
\end{figure}

\noindent At dimension six, such corrections are given by
\be
\begin{split}
\delta g_L^{b}=&-\frac 1 4 \left(\alpha_{\phi q}^{(1)}+\alpha_{\phi q}^{(3)}+\hc\right)_{33}\frac {v^2}{\Lambda^2},\\[0.1cm]
\delta g_R^{b}=&-\frac 1 4 \left(\alpha_{\phi d}^{(1)}+\hc\right)_{33}\frac {v^2}{\Lambda^2}.
\end{split}
\ee
In Figure \ref{Fig_Zbb} we show, for example, the results from the fit to an SM extension with the operators ${\cal O}_{\phi q}^{(1)}=(\phi^\dagger iD_\mu \phi)(\overline{q_L}\gamma^\mu q_L)$  and ${\cal O}_{\phi d}^{(1)}=(\phi^\dagger iD_\mu \phi)(\overline{d_R}\gamma^\mu d_R)$ (at the same time) in two cases: flavor-diagonal and family-universal interactions, and couplings only to the third family. From that figure it is obvious that bringing both observables within $1~\sigma$ of the experimental value is only possible for non-universal interactions. In that case there are two solutions allowed by EWPD: one with small deviations from both SM couplings, while the other involves a correction to $g_R^b$ so large that it actually flips the sign of the right-handed $Z\bar{b}b$ coupling.

\section{Model-independent bounds on new particles}
\label{section_MI_NP}

The use of the effective Lagrangian approach in the previous section has the advantage of being completely general. However, the proliferation of higher-dimensional operators in the expansion (\ref{EffL}) makes difficult to extract precise information of the underlying high-energy theory. The results in the last section assume only one operator at a time but, in general, a definite model will give contributions to several operators at the same time, with model-dependent correlations. Of course, it is possible to use an intermediate approach, retaining some of such correlations without giving up completely on model independence. 

Any renormalizable extension of the SM is associated with the presence of extra particles of spin 0, 1/2 or 1. Under general assumptions, all the possible particles interacting with the SM ones are determined by Lorentz and SM gauge invariance \cite{delAguila:2000rc,delAguila:2008pw,delAguila:2010mx}. Such extensions can still be studied from a model-independent point of view by considering general couplings with the SM particles. In this section we present updates of the results for the electroweak limits on extra vector-like leptons \cite{delAguila:2008pw} and new spin-1 bosons \cite{delAguila:2010mx}. 

Extra vector-like leptons, $L$, can have renormalizable interactions with the SM families only through Yukawa couplings,
$$\Delta {\cal L}=- y_{Le} \overline{L_L} \Phi_{Le} e_R- y_{Ll} \overline{L_R} \Phi_{Ll} l_L+\hc,$$
with $\Phi_{Ll, Le}$ the form of the scalar doublet needed for gauge invariance of the Yukawa terms. When the new lepton mixes with only one SM family then the mixing is given by $\left|s_{L\ell}\right|\sim \left|y_{L\ell}\right|v/\sqrt{2}M_L$. All the possible representations for new leptons that can mix with the SM ones were classified in  \cite{delAguila:2008pw}. (In the tables below we use the notation $(d_c,d_L)_Y$ to refer to the different representations of the SM gauge group, with $d_{c,L}$ the dimensionality of the color and isospin representations, and $Y$ the hypercharge.) In that reference we also computed the dimension-six effective Lagrangian obtained after integrating out the new heavy particles, and derived the resulting constraints on the lepton mixings from EWPD. Table \ref{LeptLimits} shows the updated electroweak bounds on these mixings. The most significant changes compared to the results in \cite{delAguila:2008pw}\footnote{Note that the limits in that reference are computed at the $90\%$ C.L..} occur for the case of neutrino singlets mixing with the second family. The more stringent bound in this case results from the inclusion of the unitarity constraints of the Cabibbo-Kobayashi-Maskawa matrix (first row). Because of the correlation between neutrino mixing with the electron or muon families and the Higgs mass in the fits \cite{delAguila:2008pw}, the determination of $M_H$ also has some impact in the corresponding limits.

\begin{table}[t]
\caption{Upper limit at 95 $\%$ C.L. on the absolute value of the lepton mixings, assuming each new lepton mixes with only one SM family.\label{LeptLimits}}
\centering
\begin{tabular*}{\columnwidth}{@{\extracolsep{\fill}}c l  c c c }
\ctoprule
\multicolumn{2}{c}{\!\!Lepton}&\multicolumn{3}{l}{95$\%$ C.L. EWPD limit on mixing $s_{L\ell}$}\\
~$L$&\!\!$(d_c,d_L)_Y$&~~~~~Only $e$&~~~Only $\mu$&Only $\tau$\\
\midrule
~$N$&$\left(1,1\right)_{0}$&~~~~~$0.041$&~~~$0.030$&$0.087$\\[0.4cm]
~$E$&$\left(1,1\right)_{-1}$&~~~~~$0.021$&~~~$0.030$&$0.033$\\[0.4cm]
~$\left(\!\!\!\begin{array}{c}N\\[-0.1cm]~E^-\end{array}\!\!\right)$&$\left(1,2\right)_{-\frac{1}{2}}$&~~~~~$0.020$&~~~$0.048$&$0.034$\\[0.4cm]
 ~$\left(\!\!\!\!\!\begin{array}{c}~E^-\\[-0.1cm]~~~\!E^{--}\end{array}\!\!\!\right)$&$\left(1,2\right)_{-\frac{3}{2}}$&~~~~~$0.028$&~~~$0.028$&$0.046$\\[0.25cm]
~$\left(\!\!\!\begin{array}{c}~E^+\\[-0.1cm]N\\[-0.1cm]~E^-\end{array}\!\!\right)$&$\left(1,3\right)_{0}$&~~~~~$0.019$&~~~$0.017$&$0.030$\\[0.2cm]
 ~$\left(\!\!\!\!\!\begin{array}{c}N\\[-0.1cm]~E^-\\[-0.1cm]~~E^{--}\end{array}\!\!\!\right)$&$\left(1,3\right)_{-1}$&~~~~~$0.016$&~~~$0.024$&$0.029$\\
\cbottomrule
\end{tabular*}
\end{table}

The phenomenological implications of the existence of extra vector bosons have been extensively studied in the literature. Although most studies focus only on the cases of extra neutral or charged vector bosons ($Z^\prime$ and $W^\prime$, respectively), there are many other possibilities. For the computation of the effective Lagrangian at dimension six, assuming only one new vector multiplet, all the relevant interactions between the extra spin-1 states, ${\cal V}$, and the SM particles can be written as
$$\Delta {\cal L}=-\frac{\eta_{\cal V}}{2}\left({\cal V}^{\mu~\dagger} J_\mu^{\cal V}+\hc\right),~~~J_\mu^{\cal V}=\sum_k g_{\cal V}^k j_{\mu}^{{\cal V}~\!k},$$
where $\eta_{\cal V}=1(2)$ for real (complex) vectors, $g_{\cal V}^k$ are dimensionless constants, and the dimension-three operators $j_{\mu}^{{\cal V}~\!k}$ are in the same representation of ${\cal V}$, $R_{\cal V}$. There are two kinds of non-redundant interactions: with two fermions $j^{{\cal V}\psi_1\psi_2}_\mu = [\overline{\psi_1} \otimes \gamma_\mu \psi_2]_{R_{\cal V}}$; and with two scalars and a covariant derivative $j^{{\cal V}\phi}_\mu = [\Phi^\dagger \otimes D_\mu \phi]_{R_{\cal V}}$ ($\Phi=\phi$ or $i\sigma_2 \phi^*$). The different vector representations and the corresponding currents were classified in \cite{delAguila:2010mx}, where we also compute the contributions from these particles to the effective Lagrangian at dimension six. At this order EWPD is only sensitive to the ratios $G_{\cal V}^k\equiv g_{\cal V}^k/M_{\cal V}$. Also, some of the new vectors cannot be constrained by EWPD because they only couple to quarks. In Table \ref{table_OneVLimits} we update the electroweak limits in \cite{delAguila:2010mx}, using the same assumptions in that reference. Some of these bounds were recently updated in \cite{deBlas:2012qp}. Compared to that last reference, the numbers presented here incorporate the final combination of LEP 2 $e^+ e^-\rightarrow \bar{f}f$ data. As explained above, this includes, in particular, a large set of new results in the $e^+ e^-\rightarrow \mu^+\mu^-,~\tau^+\tau^-$ channels. The good agreement of these data with the SM then strengthens the constraints on the contributions to four-lepton operators, and hence the corresponding bounds on the new vector couplings to leptons. This kind of bounds are mostly independent of the Higgs couplings to the extra vectors. 
Moreover, as explained in \cite{delAguila:2011yd}, the constraints on the contributions from new vectors to operators with four left-handed or four right-handed leptons provide the most robust limits on their couplings to fermions. Unlike the contributions to other four-fermion operator coefficients, these have a definite sign and therefore cannot be easily cancelled by other particles. More precisely, this is not possible with any kind of extra fermions or vector bosons. Relaxing the bounds derived from contributions to four-lepton interactions requires very specific scalar additions designed for this purpose \cite{delAguila:2011yd}.
  
\begin{table}[ht]
\caption{$95\%$ C.L. EWPD limits on the ratios $G_{\cal V}^k\equiv g_{\cal V}^k/M_{\cal V}$ for the extra vector bosons. For the first three representations we assume diagonal and universal fermion couplings. The results for the last six representations are obtained from a fit to each of the entries of the coupling matrices at a time.\label{table_OneVLimits}}
\centering
\begin{tabular*}{\columnwidth}{@{\extracolsep{\fill}}c c c c c}
\ctoprule
\multicolumn{2}{c}{\!\!\!\!Vector}&Parameter&\!\!$95\%$ C.L. EWPD limit\\
${\cal V}_\mu$&\!\!\!\!$(d_c,d_L)_Y$&$G_{\cal V}^k\equiv g_{\cal V}^k/M_V$&[TeV$^{-1}$]\\
\midrule
&&&\\[-0.4cm]
~~\!${\cal B}_\mu $&\!\!\!\!$\left(1,1\right)_{0}$&$G^\phi_{{\cal B}_{\phantom{G}}}$&$\left[-0.092,\phantom{+}0.092\right]$\\
                           &&$G^l_{{\cal B}_{\phantom{G}}}$&$\left[-0.123,\phantom{+}0.123\right]$\\
                           &&$G^q_{{\cal B}_{\phantom{G}}}$&-\\
                           &&$G^e_{{\cal B}_{\phantom{G}}}$&$\left[-0.157,\phantom{+}0.157\right]$\\
                           &&$G^u_{{\cal B}_{\phantom{G}}}$&-\\
                           &&$G^d_{{\cal B}_{\phantom{G}}}$&-\\
\midrule
~~\!${\cal W}_\mu $&\!\!\!\!$\left(1,3\right)_{0}$&$~G_{{\cal W}_{\phantom{G}}}^\phi$&-\\
                            &&$~G_{{\cal W}_{\phantom{G}}}^l$&$\left[-0.288,\phantom{+}0.288\right]$\\
                            &&$~G_{{\cal W}_{\phantom{G}}}^q$&-\\
\midrule
${\cal B}^{1}_\mu $&\!\!\!\!$\left(1,1\right)_{1}$&$G_{{\cal B}^1_{\phantom{G}}}^\phi$&$\left[-0.089,\phantom{+}0.089\right]$\\
                                   &&$G_{{\cal B}^1_{\phantom{G}}}^{du}$&-\\
\midrule
~~\!${\cal W}^{1}_\mu $&\!\!\!\!$\left(1,3\right)_{1}$&$|G_{{\cal W}^1}^\phi|$&$<0.301$\\
\midrule
~~\!${\cal L}_\mu $&\!\!\!\!$\left(1,2\right)_{-\frac{3}{2}}$&$|G_{\cal L}^{el}|$&$\!\!\!\!\!\!<\left(\begin{array}{c c c}0.25&0.32&0.23\\0.32&\mbox{-}&\mbox{-}\\0.23&\mbox{-}&\mbox{-}\end{array}\right)$\\
\midrule
~~\!${\cal U}^{2}_\mu $&\!\!\!\!$\left(3,1\right)_{\frac{2}{3}}$&$|G_{{\cal U}^2}^{ed}|$&$\!\!\!\!\!\!<\left(\begin{array}{c c c}0.31&0.51 &0.51\\\mbox{-}&\mbox{-}&\mbox{-}\\\mbox{-}&\mbox{-}&\mbox{-}\end{array}\right)$\\
                                &&$|G_{{\cal U}^2}^{lq}|$&$\!\!\!\!\!\!<\left(\begin{array}{c c c}0.13&0.30 &0.30\\ 0.71&0.70&\mbox{-}\\\mbox{-}&\mbox{-}&\mbox{-}\end{array}\right)$\\
\midrule
~~\!${\cal U}^{5}_\mu $&\!\!\!\!$\left(3,1\right)_{\frac{5}{3}}$&$|G_{{\cal U}^5}^{eu}|$&$\!\!\!\!\!\!<\left(\begin{array}{c c c}0.33&0.53&\mbox{-}\\\mbox{-}&\mbox{-}&\mbox{-}\\\mbox{-}&\mbox{-}&\mbox{-}\end{array}\right)$\\
\midrule
~~\!${\cal Q}^{1}_\mu $&\!\!\!\!$\left(3,2\right)_{\frac{1}{6}}$&$|G_{{\cal Q}^1}^{ul}|$&$\!\!\!\!\!\!<\left(\begin{array}{c c c}0.32&0.94&\mbox{-}\\0.61 &\mbox{-}&\mbox{-}\\\mbox{-}&\mbox{-}&\mbox{-}\end{array}\right)$\\
\midrule
~~\!${\cal Q}^{5}_\mu $&\!\!\!\!$\left(3,2\right)_{-\frac{5}{6}}$&$|G_{{\cal Q}^5}^{dl}|$&$\!\!\!\!\!\!<\left(\begin{array}{c c c}0.98&1.24 &\mbox{-} \\0.96 &\mbox{-} &\mbox{-} \\0.96 &\mbox{-} &\mbox{-} \end{array}\right)$\\
                                   &&$|G_{{\cal Q}^5}^{eq}|$&$\!\!\!\!\!\!<\left(\begin{array}{c c c}0.71&0.79\phantom{0}&1.10\phantom{0} \\\mbox{-}&\mbox{-}&\mbox{-}\\\mbox{-}&\mbox{-}&\mbox{-}\end{array}\!\!\!\!\right)$\\
\midrule
~~\!${\cal X}_\mu $&\!\!\!\!$\left(3,3\right)_{\frac{2}{3}}$&$|G_{\cal X}^{lq}|$&$\!\!\!\!\!\!<\left(\begin{array}{c c c}0.22&0.77 &0.61 \\1.23 &1.53&\mbox{-}\\\mbox{-}&\mbox{-}&\mbox{-}\end{array}\right)$\\
\cbottomrule
\end{tabular*}
\end{table}


\section{Conclusions}
\label{conclusions}

Electroweak precision data has been crucial in testing the validity of the SM at the electroweak scale. The model predictions have been proved to the level of radiative corrections, finding a good agreement with data. 
This, in turn, implies strong constraints on any possible contribution from new physics to electroweak precision observables. In this short note we have studied the EWPD constraints on new physics from a model-independent point of view, using an analysis based on a dimension-six effective Lagrangian. EWPD provides strong limits on many of the different dimension-six operators. We have presented updated results for a large set of these interactions with the most up-to-date data. We have also updated the limits for several different SM extensions including extra leptons and vector bosons. 

The results of these EWPD analyses of new physics guide and provide complementary information to that from direct searches. With about 20 fb$^{-1}$ of data collected at the LHC at $\sqrt{s}=8$ TeV we still do not have any significant deviation from the SM, so we can use the LHC results to derive indirect constraints. With the current precision of data, the electroweak bounds are comparable to those from the LHC at 8 TeV, and in many cases still dominate \cite{deBlas:2013qqa}. However, instead of focusing on the competition between EWPD and LHC limits, we would like to emphasize the complementarity of both, and the importance of a combined analysis. The interest of this kind of combination is two-fold. On the one hand, for those interactions that can contribute to both types of observables (electroweak and LHC), this allows to strengthen current constraints \cite{deBlas:2013qqa}. On the other hand, given the different nature of the LHC and electroweak observables, both can be sensitive to certain interactions that cannot be seen by the other. For instance, EWPD are virtually blind to four-quark contact interactions, which can be tested at the LHC \cite{Domenech:2012ai}. Even if one of the experiments is not sensitive to some interaction, the combined limits can be significantly stronger within definite scenarios because of theoretical correlations. Thus, if we aim to cover the most general class of new physics scenarios, and obtain the most stringent bounds, the combination of electroweak and direct searches analyses becomes mandatory.  


\section*{Acknowledgements}
It is a pleasure to thank F. del \'Aguila and M. P\'erez-Victoria for collaboration in this topic. This work has been supported in part by the U.S. National Science Foundation under Grant PHY-1215979.

%

\end{document}